\documentclass[doublecol]{epl2}
\usepackage{color}
\usepackage{graphicx}
\usepackage{amssymb}
\usepackage{amsmath}
\usepackage{amsfonts}
\usepackage[normalem]{ulem}
\usepackage{times}
\usepackage{float}
\usepackage{epstopdf}
\usepackage{subcaption}

\newcommand{\Tc}{T_{\text c}}

\newcommand{\gammadot}{\dot{\gamma}}

\newcommand{\tauLJ}{\tau_\mathrm{LJ}}
\newcommand{\taurelax}{\tau_\mathrm{relax}}

\newcommand{\rhoA}{\rho_{\text A}}
\newcommand{\rhoB}{\rho_{\text B}}

\newcommand{\myeq}{\!=\!}
\renewcommand{\vec}{\mathbf}

\newcommand{\sigmaAA}{d_{\mathrm{AA}}}

\newcommand{\rcab}{r_{\mathrm{c},\alpha\beta}}

\newcommand{\Uwall}{U_{\mathrm{wall}}}

\newcommand{\Cexz}{\ensuremath{C_{\varepsilon_{xz}}}}
\newcommand{\exz}{\ensuremath{\varepsilon_{xz}}}
\newcommand{\exzna}{\ensuremath{\varepsilon^\text{n.a.}_{xz}}}
\newcommand{\rpar}{\ensuremath{r_{\parallel}}}

\newcommand{\delete}[1]{}

\newcommand{\blue}[1]{\textcolor{black}{#1}}

\newcommand{\hide}[1]{\textcolor{blue}{}}

\usepackage{hyperref}

\title{Wall effects on spatial correlations of \blue{non-affine} strain in a 3D model glass}
\author{M. Hassani\inst{1}, P. Engels\inst{1}, F. Varnik\inst{1}
\thanks{Corresponding author: \email {fathollah.varnik@rub.de}}}
 \shortauthor{Hassani \etal}
 \shorttitle{Wall effect on correlations of plastic events}
 \institute{
   \inst{1} ICAMS, Ruhr-Universit\"at Bochum, Universit\"atsstra{\ss}e 150, 
44780 Bochum, Germany
}
\abstract{
Effects of hard planar walls with a particle scale roughness on the spatial correlations of \blue{non-affine} strain in amorphous solids are investigated via molecular dynamics simulations. When determined within layers parallel to the wall plane, normalized \blue{non-affine} strain correlations are enhanced within layers closer to the wall. The amplitude of these correlations, on the other hand, is found to be suppressed by the wall. While the former is connected to the effects of a hard boundary on the continuum mechanics scale, the latter is attributed to molecular scale wall effects on the size of the region (nearest neighbor cage), explored by particles on intermediate times scales.
}

\begin{document}

\maketitle

\section{Introduction}

The nature of plasticity in amorphous solids has been the subject of intense research in recent years \cite{Argon1979,Barrat2011}. Unlike crystalline materials  which yield upon loading along well defined glide planes, the lack of long range order makes the prediction of the plastic deformation in amorphous materials a challenging task. A promising research direction has been the study of spatio-temporal behavior of irreversible particle rearrangements~\cite{Lemaitre2009, Martens2011, Hassani2016}. The energy accumulation by the loading, beyond the yielding strain, is frequently interrupted by sudden drops due to the localized plastic rearrangements \cite{Dasgupta2012,Varnik2004,Varnik2004a,Hassani2016}. \blue{These irreversible structural changes, on the other hand, are often preceded by localization of non-affine deformation~\cite{Tsamados2008,Ketov2015}}.

In close connection with these observations, there is growing agreement on the role of elasticity for mediating correlations of fluctuations in amorphous solids~\cite{Argon1979,Picard2004,Picard2005,Bocquet2009}. As suggested by Argon in his seminar paper~\cite{Argon1979}, the strain associated with a plastic event generates, much in the same way as in the Eshelby's inclusion problem ~\cite{Eshelby1957}, a long-range \emph{elastic} strain field in the surrounding medium. \blue{Due to the heterogeneity of local elastic properties~\cite{Barrat2011}, this may give rise to significant strain fluctuations far from the original event.}

\blue{In view of the close connection between local no-affine strain and plastic activity, a study of spatial correlations of non-affine strain} may provide useful information on the underlying mechanisms of mechanical deformation in amorphous solids. Important contributions have been made by experiments and computer simulations, revealing avalanche-like plastic response 
~\cite{Lemaitre2009,Chattoraj2011,Lemaitre2014} and long-range spatial 
correlations between elementary plastic 
events~\cite{Maloney2006a,Maloney2009,Chikkadi2011,Chikkadi2012b,Mandal2013,Varnik2014,Nicolas2014a}. Recently, non-equilibrium mode-coupling theory~\cite{Fuchs2003a} has been extended to study spatio-temporal correlations of non-affine strain~\cite{Illing2016}, providing a theoretical basis for the newly observed Eshelby-type strain correlations in supercooled liquids~\cite{Chattoraj2013}.

While most of these studies have focused on the mechanical response in the bulk, effects arising from boundaries have found less attention~\cite{Picard2004,Nicolas2013}. However, in the past twenty years, it has been well established that confining walls may have a significant effect on the dynamics of glass formers~\cite{Baschnagel2005}. Both experiments and numerical simulations have shown that, while smooth walls enhance the dynamics of structural relaxation~\cite{Fehr1995,Anastasiadis2000,Varnik2002c,Varnik2002e,Keddie1994a}, the relaxation time increases by orders of magnitude in the vicinity of substrates with a roughness on the length scale of a particle diameter~\cite{Ellison2002,Scheidler2004,Baschnagel2005,Varnik2009}. Moreover, when viewed from the perspective of continuum elasticity, a hard wall may significantly alter the response of a solid medium to a locally imposed strain. This change of the elastic propagator, would then imply that the effect of a relaxation event on triggering further events is influenced by a hard wall.

In this letter, we address this issue via molecular dynamics (MD)  simulations of a model glass in three dimensions. The main focus of the work is on the effect of walls on spatial correlations of \blue{non-affine} strain under steady shear. Strain-strain correlations are evaluated within thin layers parallel to the walls. It is found that the wall enhances spatial correlations of non-affine deformation within the planes close to the wall. Away from the wall, the decay occurs faster, approaching the bulk behavior. \blue{This observation is corroborated by continuum mechanics (CM) results for the strain field around a pre-sheared spherical inclusion, placed in a homogeneous and isotropic solid, at various distances from a hard wall. MD studies of this inclusion problem support further the close connection between correlations of non-affine strain and continuum level response.} Interestingly, the continuum scale enhancement of correlations is accompanied by a dramatic suppression of the strain amplitude, the latter attributed to molecular scale effects.

\section{Method}
\label{sec:method}

\subsection{Model} \label{sec:model}
For the MD part of this study, we use the well known Kob-Andersen binary (80:20) Lennard-Jones glass \cite{Kob1994} 
at a total density of $\rho\myeq \rhoA+\rhoB \myeq 1.2$. At this number density, 
the mode coupling critical temperature (a measure of the proximity to the glass 
transition) of the model is $\Tc \myeq 0.435$ \cite{Kob1995a}. A and B particles 
in the model interact via $U_{\mathrm{LJ}}(r)\myeq 
4\epsilon_{\alpha\beta}[(d_{\alpha\beta}/r)^{12}-(d_{\alpha\beta}/r)^6],$ with 
$\alpha,\beta\myeq {\mathrm{A,B}}$, $\epsilon_{\mathrm{AB}}\myeq 
1.5\epsilon_{\mathrm{AA}}$, $\epsilon_{\mathrm{BB}}\myeq 
0.5\epsilon_{\mathrm{AA}}$, $d_{\mathrm{AB}}\myeq 0.8d_{\mathrm{AA}}$, 
$d_{\mathrm{BB}}\myeq 0.88d_{\mathrm{AA}}$ and $m_{\mathrm{B}}\myeq 
m_{\mathrm{A}}$. In order to enhance computational efficiency, the potential is 
truncated at twice the minimum position of the LJ potential, $\rcab\myeq 2.245 
d_{\alpha\beta}$. The parameters $\epsilon_{\mathrm{AA}}$, $d_{\mathrm{AA}}$ and 
$m_{\mathrm{A}}$ define the units of energy, length and mass, respectively.  The unit of time is a combination of these 
units, $\tauLJ \myeq d_{\mathrm{AA}}\sqrt{m_{\mathrm{A}} / \epsilon_{\mathrm{AA}}}$. Unless otherwise stated, the simulation box is a 
cube of length $L=100$, containing $1.2\times10^6$ particles. All the simulations reported here are performed using 
LAMMPS~\cite{Plimpton1995} with a discrete time step of $dt \myeq 0.005$.

The model has been investigated in previous works, addressing various issues such as non-Newtonian rheology \cite{Berthier2002a,Varnik2006d}, heterogeneous plastic deformation and flow \cite{Varnik2003,Hassani2016} and structural relaxation under shear \cite{Berthier2002a,Varnik2006b}.

%\label{sec:Sample preparation}Sample preparation
\subsection{Sample preparation} Five statistically independent samples are prepared by equilibrating the system at a high temperature of $T\myeq 1$ (liquid state) and then quenching to a temperature of $T=0.2$ (glassy state). The system is then aged for $ t_\text{w} \myeq 4\times10^4\tauLJ$. This ensures that the structural relaxation time in the glassy state obeys $\taurelax\ge 10^4\tauLJ$~\cite{Varnik2004}. This time is large compared to the time intervals of interest here. Thus, the samples prepared at $T=0.2$ are expected to behave like an amorphous solid for all the quantities of interest in this study.

\subsection{Imposing shear} Shear is imposed in the $xz$-plane by moving two planar and parallel walls with a relative velocity of $2\Uwall$. Given the separation, $L_z$, between the walls, this defines the overall shear rate $\gammadot=2\Uwall/L_z$. The walls are prepared by freezing two layers of particles on the bottom and upper parts of the simulation cell, each with a thickness  of three particle diameters. This is larger than the present cutoff range of the LJ potential, so that LJ-particles on both sides of a layer do not interact with each other. Wall particles in the present set of simulations have no thermal motion but move all together with a constant velocity of $\pm \Uwall$ along the $x$-direction.

\begin{figure}
\begin{center}
\includegraphics[width=8cm]{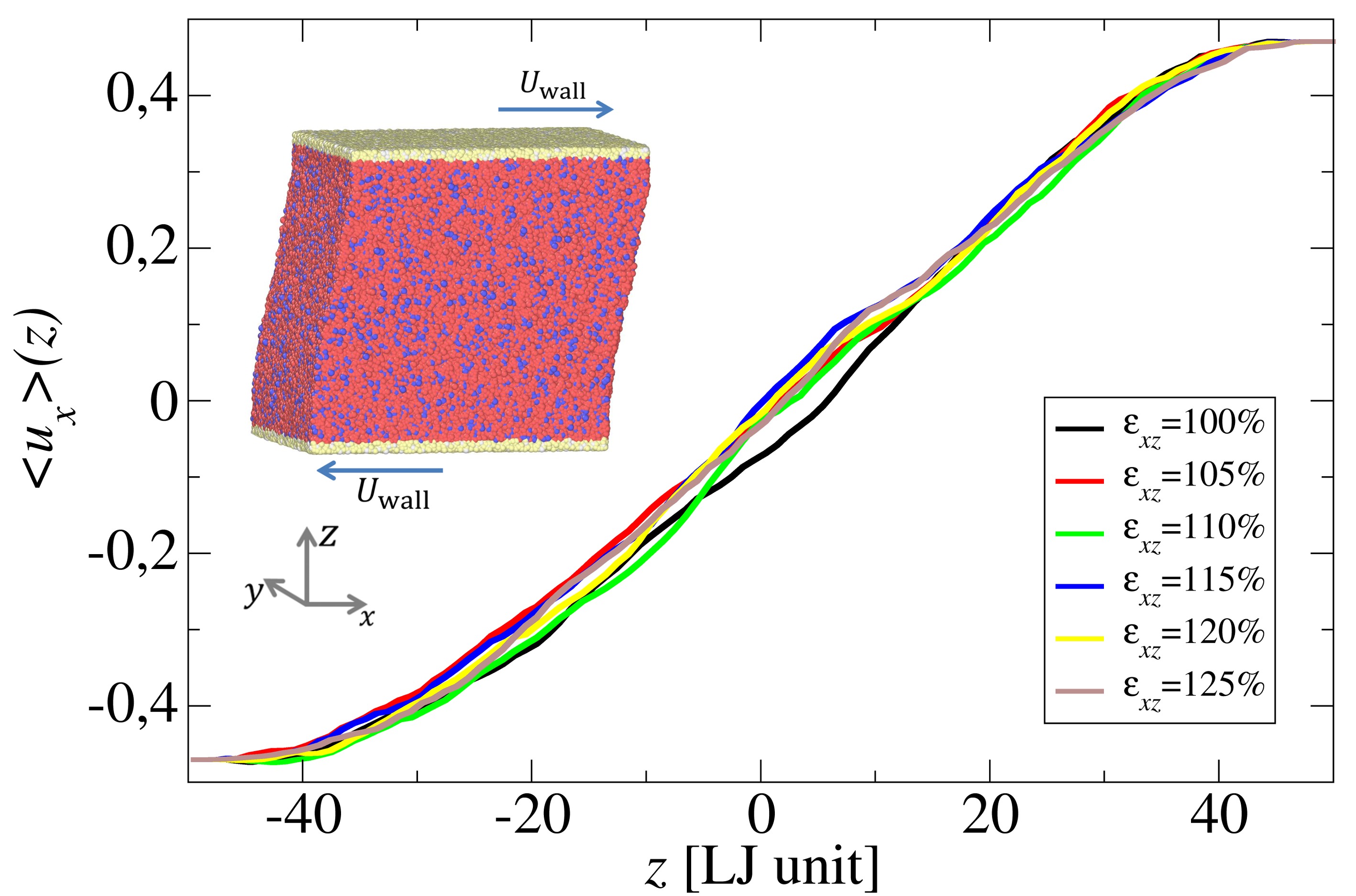}

\end{center}
\caption[]{The displacement profiles, $u_x(z)$, in the steady shear regime. The position $z=0$ corresponds to the midway between the walls. For each $\exz$ indicated in the plot, the corresponding $u_x(z)$ is evaluated in a strain interval of 1\%, from $\exz$ to $\exz+1$\%. As can be seen from this plot, at such a short strain interval, the displacement profile is essentially linear. The inset shows a snapshot of the simulation cell  with A and B particles shown in red and blue colors, respectively. The system is deformed by the relative motion of the solid walls (gold color) with the velocity $\pm \Uwall$.}
\label{fig:snapshot}
\end{figure}

\section{Strain correlations under steady shear}
In order to access correlations of non-affine strain, we start with particle trajectories, $\vec{r}_i(t)$ ($i$ is the particle index), and the corresponding displacement vectors over the time interval $\delta t$, $\vec{u}_i(\delta t)=\vec{r}_i(t+\delta t)-\vec{r}_i(t)$. We note that the time $t$ is irrelevant here due to the steady shear condition (average over $t$ helps to improve statistical accuracy, though). For a layer of thickness $\Delta z$ with centre at $z$, the non-affine part of the displacement vector is obtained by subtracting the contribution of the globally imposed (affine) deformation via  $\vec u^\text{n.a.}(\vec r_i, \delta t)=\vec{u}_{i}-\left\langle\vec{u}\right\rangle_\text{layer}(z)$. Here, $\left\langle \vec u \right\rangle_\text{layer}(z)$ is the average displacement vector of all the particles within that layer (i.e., with $z_i$ satisfying $z - \Delta z/2 \le z_i \le z+\Delta z/2$). In order to reduce the statistical noise, the thus defined non-affine displacement field is averaged over a length scale, $w$, usually of the order of the nearest neighbor distance. This coarse-graining process is performed via 
$\vec{u}^\text{CG}(\vec{r},\delta t)={\sum_{i}{\vec{u}^\text{n.a.}}(\vec r_i, \delta t)\phi(||\vec r-\vec r_i||)}/{\sum_{j}\phi(||\vec r-\vec r_j||)}$, using the coarse-graining function, 
$\phi(r)=\frac{1}{\pi\omega^2}e^{-(r/w)^2}$ \cite{Goldenberg2007}. Here, $r=||\vec r||=\sqrt{x^2+y^2+z^2}$ is the magnitude of the three dimensional vector $\vec r$, meaning that the averaging is performed within a sphere of radius $\sim w$. 

In the following, we will drop the argument $\delta t$ but will keep in mind that the displacement field and all the quantities derived from it are obtained within a finite time interval, $\delta t$. Using the thus defined coarse-grained non-affine displacement field, the corresponding strain field and its spatial correlation are obtained as $\boldsymbol\varepsilon^\text{n.a.}(\vec{r})=(\nabla \vec{u}^\text{CG}(\vec{r})+(\nabla \vec{u}^\text{CG}(\vec{r}))^T)/2$, and $\Cexz(\vec{r})= \langle \exzna(\vec{r} + \vec{r}_0)  \exzna(\vec{r}_0)\rangle$. We have evaluated the gradients of the coarse-grained deformation field, $\vec{u}^\text{CG}(\vec{r})$, both via finite differencing and by invoking gradients of $\phi$. Essentially the same results are obtained within both approaches. Here we report the data using the second approach. \blue{We have also examined via comparisons with non-coarse-grained data that the use of coarse-graining procedure does not bias the spatial dependence of the correlation function but only reduces the noise level (not shown).}

\begin{figure}
\begin{center}
\includegraphics[width=8cm]{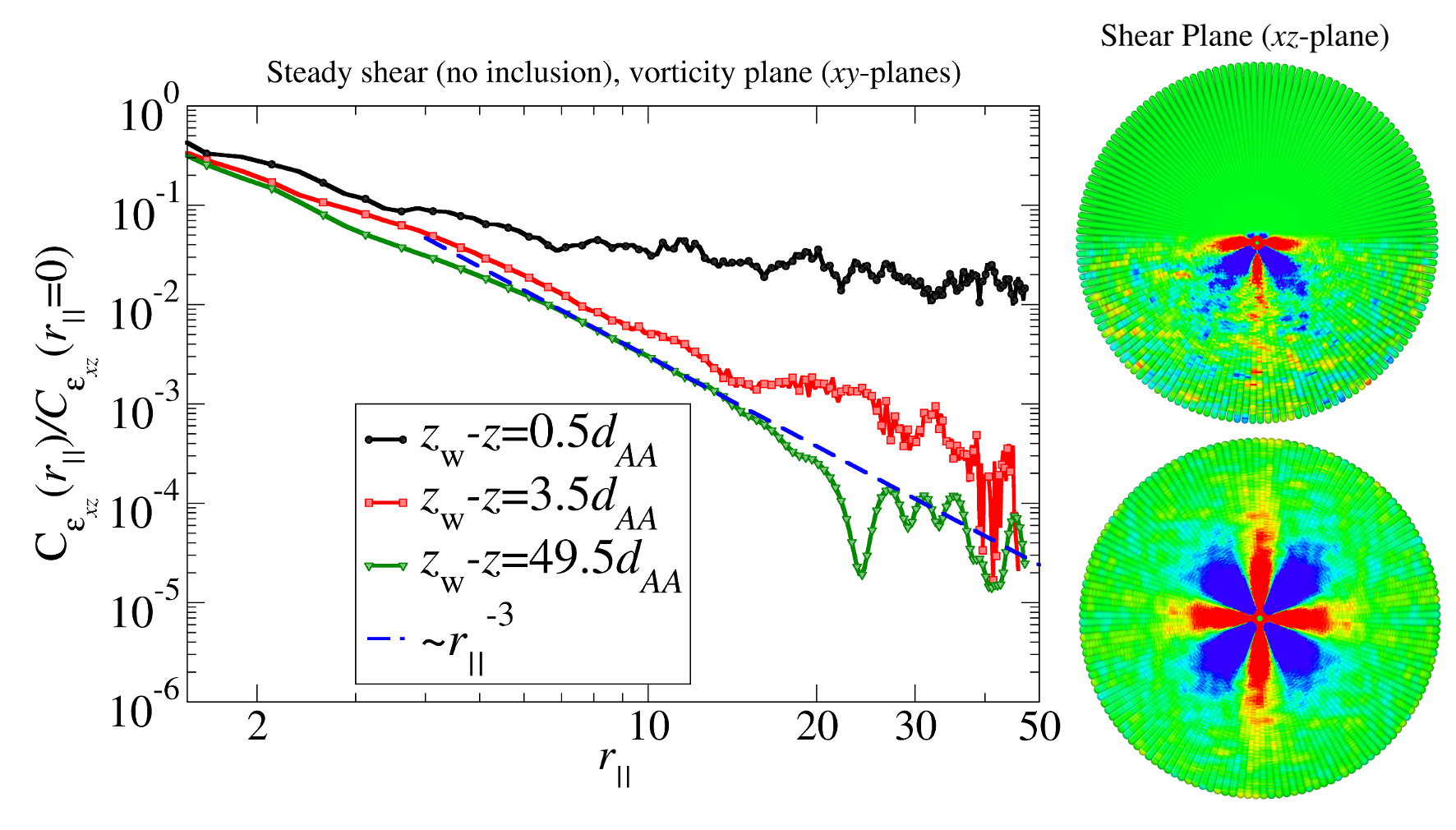}
\end{center}
\caption[]{Normalized correlations of non-affine shear strain, $\Cexz(\rpar,z) / \Cexz(0,z)$, versus distance, $\rpar=\sqrt{x^2+y^2}$, between two rearrangement events. $\Cexz$ is evaluated within thin parallel layers with a thickness equal to one particle diameter. The coordinate $z_\text{w}$ gives the wall position and $z_\text{w}-z$ the distance of the layer's centre to the wall. The curves give $\Cexz$ after integration over the polar angle within the $xy$-plane. An  enhancement of the normalized correlations (a slower decay) within layers closer to the wall is clearly visible. Moreover, the correlation function within a layer away from the wall approaches $1/\rpar^3$ decay for large $\rpar$ (dashed line). 
The color maps on the right show the non-affine strain correlations within the $xz$-plane, highlighting the broken symmetry close to the wall (upper image) and the 4-fold symmetry far away from it (lower image).}
\label{fig:wallDriven}
\end{figure}

The strain field and the spatial correlations thereof are calculated in the steady state, after neglecting the first $100\%$ strain. Unless otherwise stated, the time interval, $\delta t$, for the evaluation of displacements corresponds to $1\%$ strain. Figure~\ref{fig:wallDriven}  illustrates the thus computed correlations of non-affine shear strain. $\Cexz$ has been evaluated within thin layers parallel to the wall ($xy$-plane). The layer thickness is equal to one particle diameter. Since each layer has a well-defined distance, $z$, from the wall, one can examine whether and how the strain correlation function changes with distance to the wall. In order to improve statistical accuracy for a quantitative survey, the correlation function is averaged over the polar angle within each layer. The result is shown in the main plot in Fig.~\ref{fig:wallDriven}. As seen from this data, the normalized strain correlations, $\Cexz(\rpar,z)/\Cexz(0,z)$, decay more slowly within layers closer to the wall ($\rpar$ stands for the distance between two events within the same $xy$-plane, $\rpar=\sqrt{x^2+y^2}$). Hence, the presence of a hard solid wall enhances the mutual influence of local relaxation events at two distant points. Taking this analogy to the continuum mechanics level, this would mean that the elastic propagator, which connects any two rearrangement events in the system is enhanced in the presence of a hard wall. This interpretation is supported by the results presented in the next section.

\section{Pre-sheared inclusion close to a wall}
\blue{The effect wall on elastic propagator is addressed on the continuum scale level. For this purpose,} spectral solver of our house-made software OpenPhase (\url{www.openphase.de}) is used for numerical solution of the continuum mechanics equations. Since the spectral solver uses fast Fourier transform to solve the equations of elasticity, it requires periodic boundaries in all spatial directions. This is accounted for by a periodic system consisting of two elastic domains, (i) a first domain, with an elastic modulus of $E_1=70\times 10^9$ Pascal and a Poisson ratio of $\nu=0.25$, where we embed a strained inclusion of radius $a$ in a medium of size $L\approx25a$ with the 
same properties as the matrix and (ii) a second domain with orders of magnitude higher elastic modulus ($E_2=100E_1$) and the same Poisson ratio to mimic a hard wall (study of the ratio of the two elastic moduli is also performed to make sure the result is independent of this ratio, not shown here). The periodic image of the upper wall acts as a hard lower boundary. The equation of mechanical equilibrium, $\vec{\nabla} \cdot \boldsymbol{\sigma} = \mathbf{0}$, where $\boldsymbol{\sigma}$ is the stress tensor, is solved for an initial strain of $\epsilon^{*}_{xz}=0.01$ in the inclusion and zero elsewhere.

\begin{figure}
	\begin{center}
		\includegraphics[width=8.8cm]{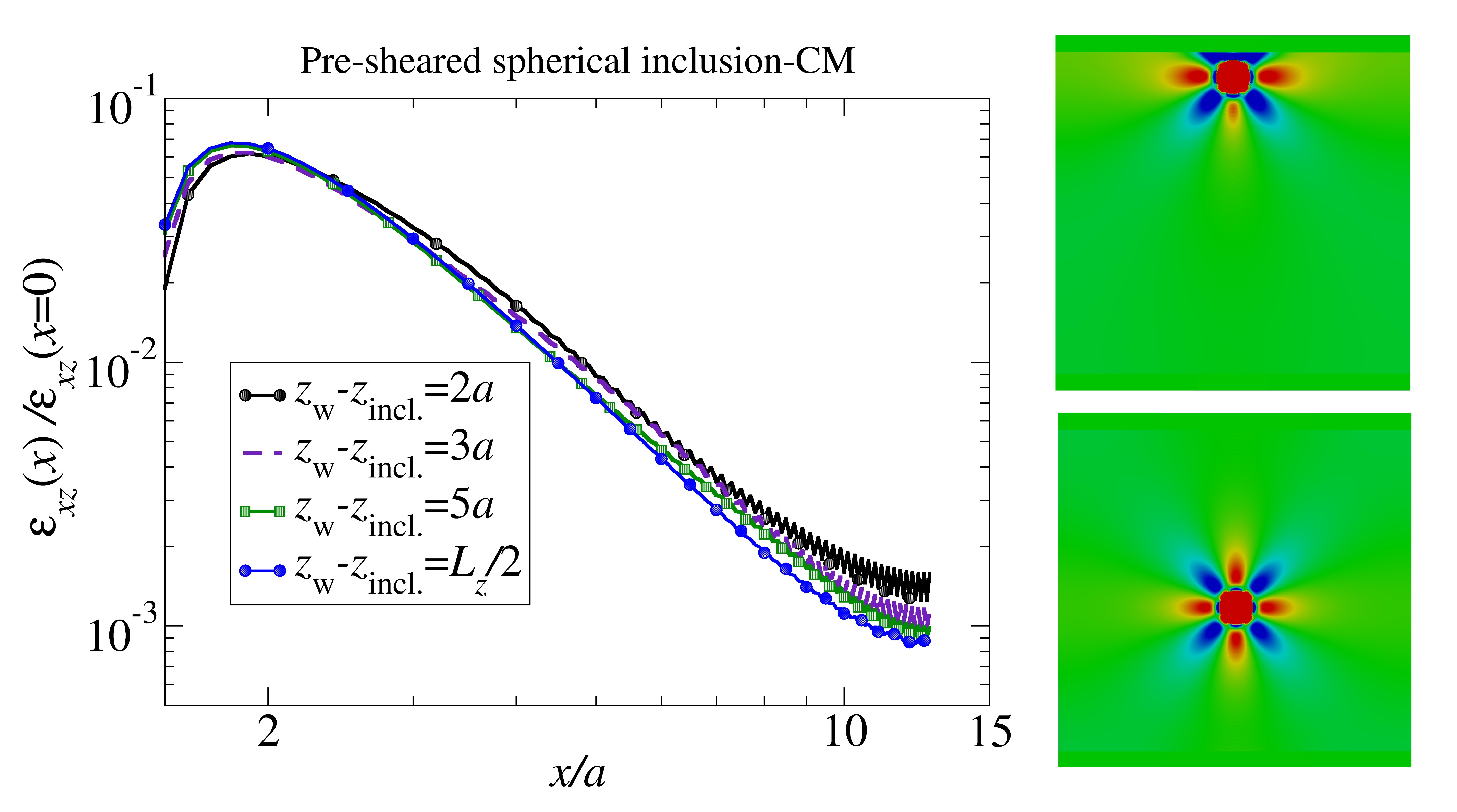}
	\end{center}
	\caption{Normalized shear strain profile, $\left\langle\exz\right\rangle(x)/\left\langle\exz\right\rangle (x\myeq0)$ along a line parallel to the $x$-axis, which passes through the center of a spherical inclusion. The profiles are displayed for different wall-inclusion distances, $z_\text{w}-z_\text{incl.}$, as indicated. $z_\text{w}$ is the position of the wall and $z_\text{incl.}$ refers to the $z$-coordinate of the inclusion's centre. Results are from numerical continuum mechanics calculations. Strain profiles show a slower decay when the 
		inclusion is closer to the wall. Distances are normalized by the radius of inclusion, $a$. The color maps on the right show the strain distribution within the $xz$-plane, highlighting the broken symmetry for an inclusion close to the wall (upper image) and the 4-fold symmetry far away from it (lower image).}
	\label{fig:CM-MD-EEI}
\end{figure}

\blue{Results of these continuum scale calculations are shown in Fig.~\ref{fig:CM-MD-EEI} and reveal, in a way similar to correlations of non-affine strain,  an enhancement of the strain field around an inclusion at closer distances to the wall. The close connection between local relaxation events in amorphous solids and the embedded inclusion problem~\cite{Eshelby1957} has proved to be quite useful in interpreting spatial strain correlations in the bulk (see, e.g.,~\cite{Talamali2012,Dasgupta2013,Nicolas2015} and references therein).} Results presented here clearly underline that this analogy persists in the presence of hard boundaries. We remark that an enhancement of the propagator by a hard wall can be also inferred from~\cite{Nicolas2013} but the connection to correlations of local strain fluctuations has not been explored there. \blue{Moreover, the problem considered here is three dimensional, whereas the solution provided in ~\cite{Nicolas2013} addresses the two dimensional case.}

\begin{figure}
\begin{center}
	\includegraphics[width=8cm]{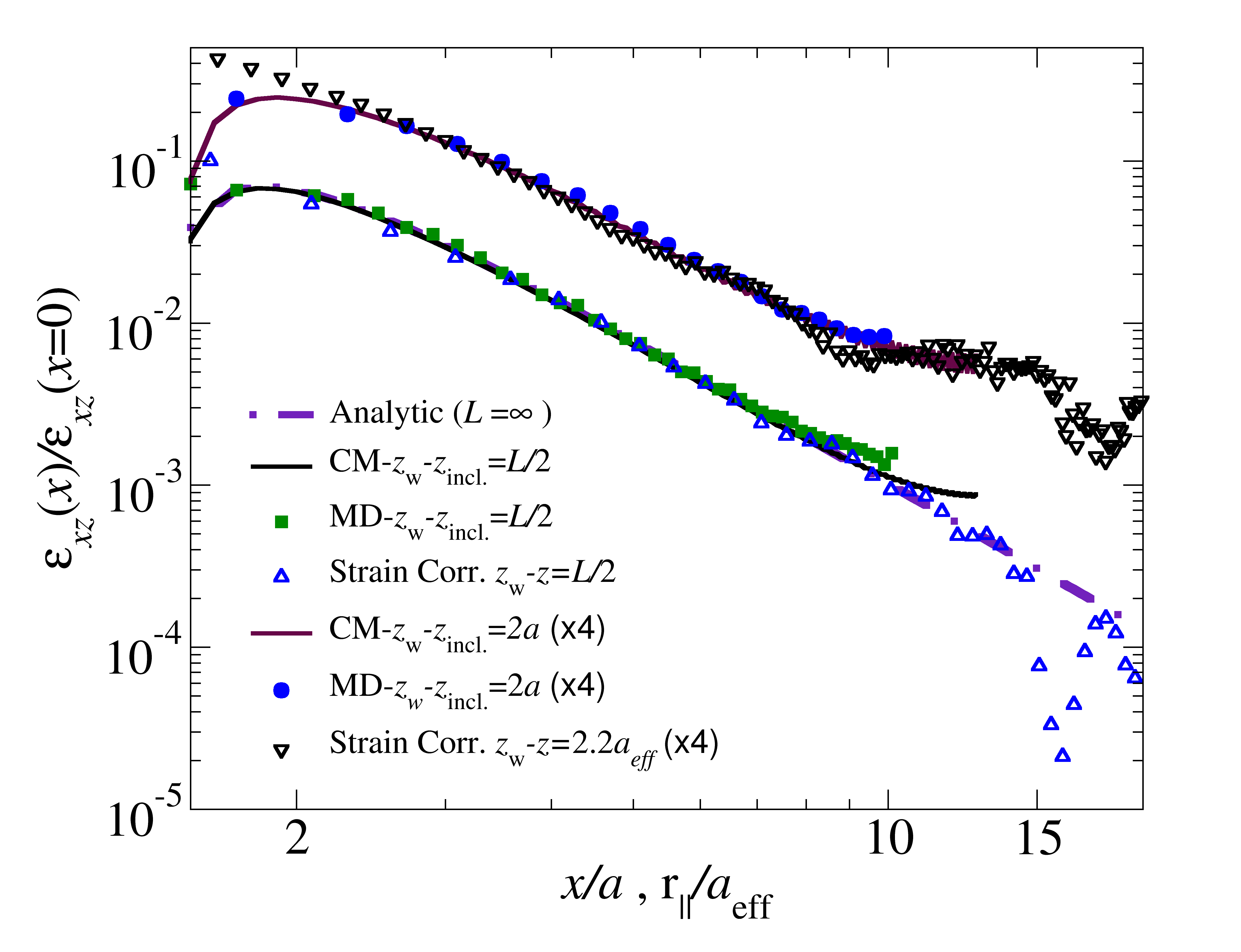}
\end{center}
\caption{\blue{Comparison of continuum mechanics results on normalized shear strain profile induced by a pre-sheared spherical inclusion (solid lines) with molecular dynamics data on spatial correlations of non-affine strain (open triangles). The CM data are plotted versus the normalized distance $x/a$ ($a$ is the inclusion radius) from the center of inclusion. The correlation function is plotted versus $r_\parallel/a_\text{eff}$, where $r_\parallel$ is the distance between two point within a parallel layer and $a_\text{eff}\approx 1.6\sigmaAA$ is the effective size of an elementary shear event (shear transformation zone). $z_\text{w}-z$ gives the distance from the wall. For clarity, part of the data is shifted upwards via multiplication by a factor of 4 (note the log-log scale). The analytic solution for the strain around an inclusion in an unbounded system~\cite{Dasgupta2013} confirms the validity of CM calculations away from the wall. As a further validation of CM calculations, MD results for the case of inclusion are also shown (filled symbols). The most striking result here is, however, the agreement between the behavior of the non-affine \emph{strain correlations} and the Eshelby-type strain field caused by an inclusion.}}
\label{fig:compareWithAnal} 
\end{figure}

\blue{For a quantitative comparison of the results for the inclusion problem and the correlation of strain fluctuations, one must first specify the size of a rearranging region which would play the role of an "inclusion". This is expected to be comparable to the size of the cage around a reference particle but its exact numerical value is unknown. As shown in	Fig.~\ref{fig:compareWithAnal}, we find that a value of $a_\text{eff}\approx 1.6\sigmaAA$ yields a good agreement between the thus rescaled non-affine strain correlations and the CM result for the strain field around a pre-sheared spherical inclusion both in the proximity of a hard wall as well as in the bulk. In this latter case, analytic solution of the Eshelby inclusion problem is available and supports the validity of the numerical results. Noteworthy, the curve in Fig.~\ref{fig:wallDriven} corresponding to $z=0.5\sigmaAA \approx 0.3a_\text{eff}$ has no analog in CM calculations, where the center of inclusion cannot come closer to the wall than its radius.}

\blue{To validate the continuum scale result in the proximity of the wall, the inclusion problem is solved also via MD simulations by applying the procedure described in ~\cite{Puosi2014}. As shown in Fig.~\ref{fig:compareWithAnal}, results of these simulations support the validity of the CM data both close to and away from the wall.}

\section{Strain amplitude}
\begin{figure}
\begin{center}
\includegraphics[width=8cm]{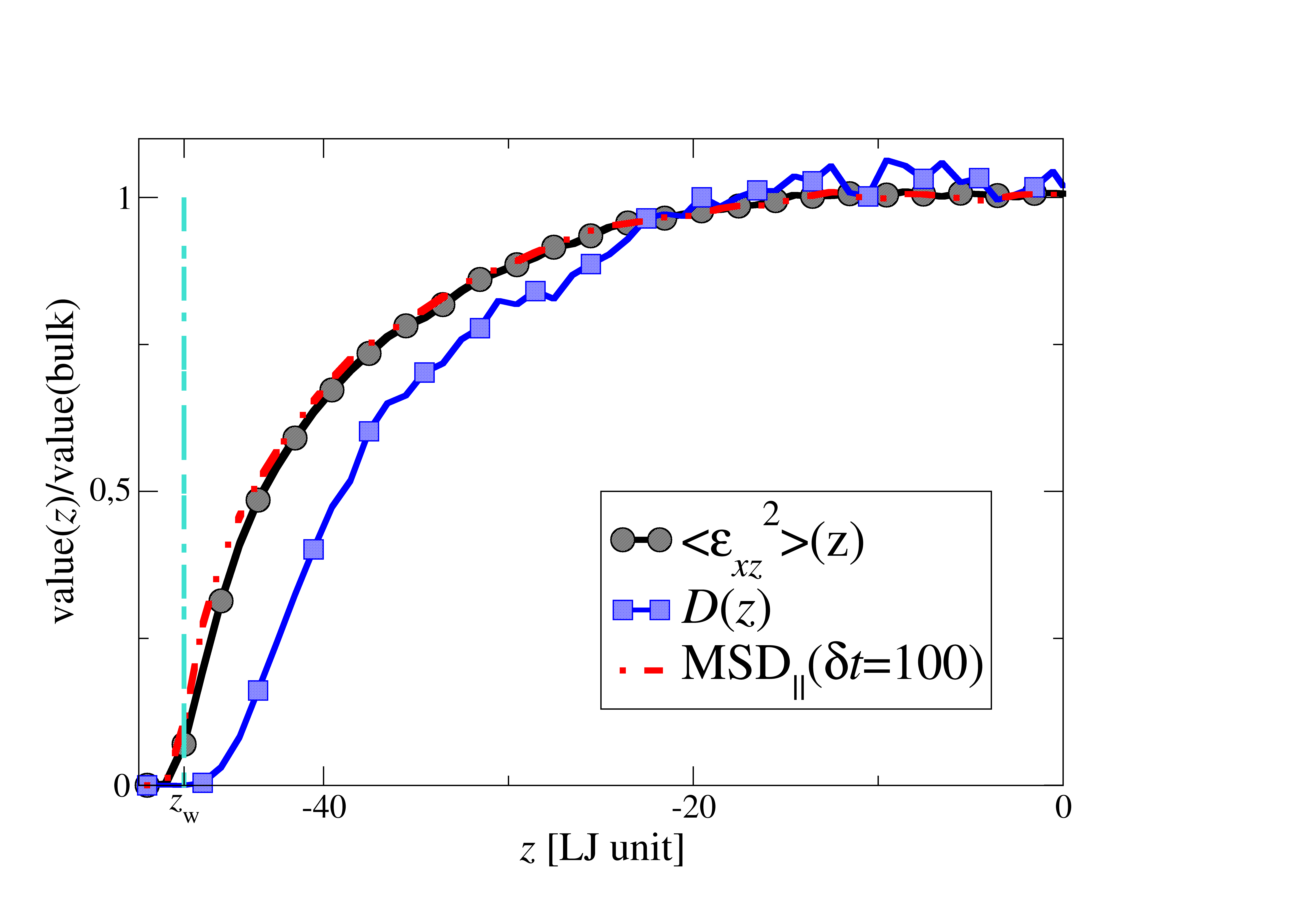}
\includegraphics[width=8cm]{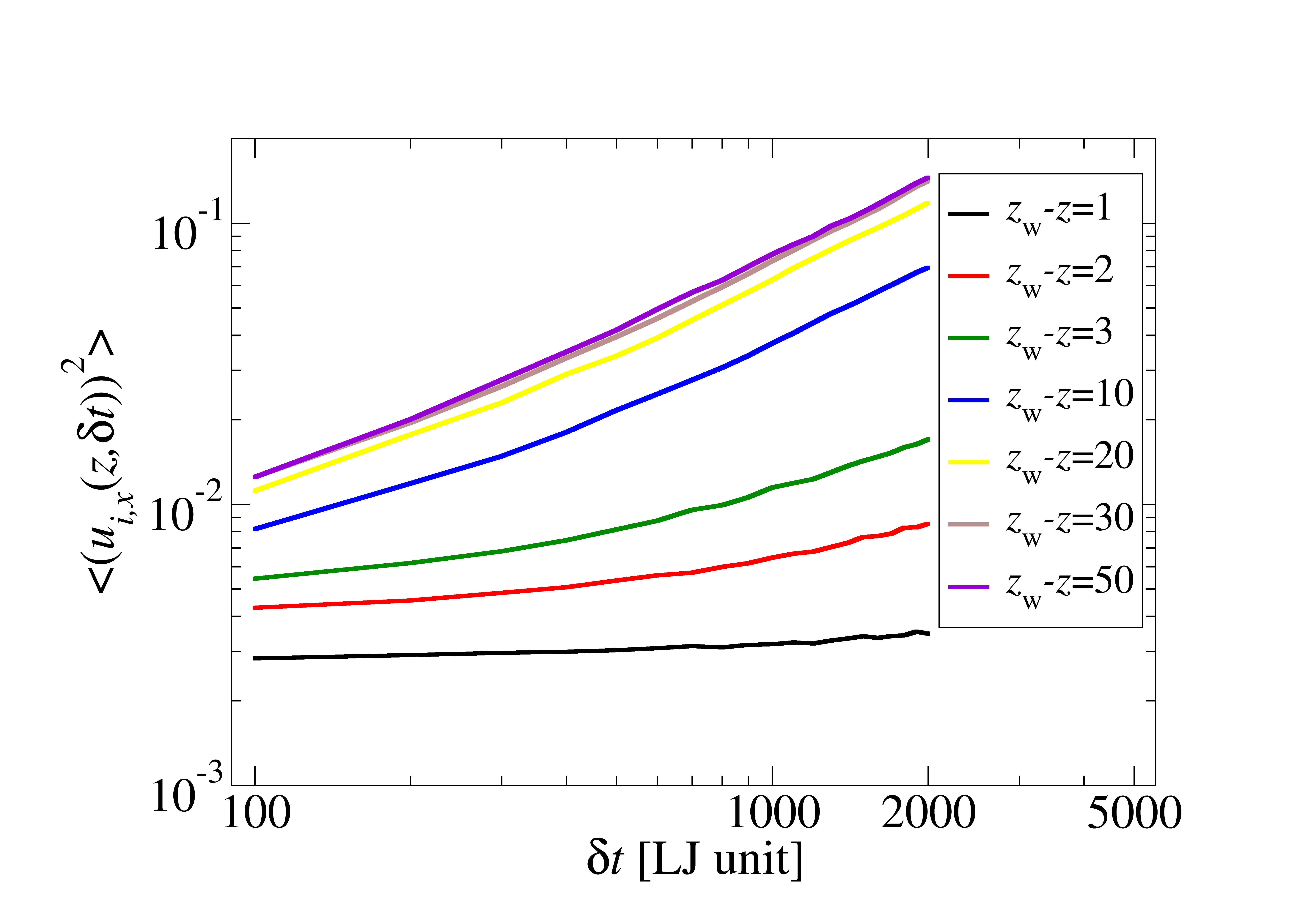}
\end{center}
\caption[]{(a) Wall effect on the non-affine strain amplitude, $\langle \exz^2 \rangle = \Cexz(0,z)$,
and the layer-resolved diffusion coefficient, $D(z)$. 
Both $D(z)$ and $\Cexz(0,z)$ decay when approaching the wall $(z\to z_{\text{w}})$. The effect of wall on diffusion coefficient, however, seem to be significant at larger distances.
(b) Layer-resolved mean squared displacement along a direction parallel to the wall (MSD$_\parallel$) versus time, $\delta t$. The smallest recorded time is $\delta t=100$. This corresponds to a globally imposed strain interval of 1\% (shear rate, $\gammadot=10^{-4}$), within which all reported non-affine strain correlations and the strain-amplitude are evaluated.}
\label{fig:strAmpl}
\end{figure}

In addition to the spatial correlations of non-affine strain, walls may also have strong influence on the strength of single \blue{rearrangement} events. Indeed, previous studies of diffusion and structural relaxation both in equilibrium~\cite{Ellison2002,Varnik2002e,Scheidler2004,Baschnagel2005,Varnik2009} and under shear~\cite{Varnik2002d} show that rough walls do in general slow down the rearrangement dynamics. One can thus expect a decrease of the strain amplitude in the proximity of the wall. This expectation is born out by the results shown in Fig.~\ref*{fig:strAmpl}a, where the layer-resolved mean squared non-affine strain, $\langle \exz^2 \rangle\equiv \Cexz(0,z)$, is shown versus the layer-distance from the wall. The data is normalized with respect to the bulk behavior thus directly revealing wall effects. It is clearly seen that the strain amplitude is suppressed in the vicinity of the wall. Interestingly, however, a plot of the normalized layer-resolved shear-induced diffusion coefficient, $D(z)/D^\text{bulk}$, reveals that wall effects on single particle diffusion are "delayed" by a distance of the order of a few particle diameters.

We have examined the layer-resolved structural relaxation time and observe the same trend (not shown). This dissimilarity is resolved by realizing that, in contrast to diffusion coefficient, which reflects the particle dynamics on long time scales, where structural relaxation has taken place completely, the non-affine strain is evaluated within a time interval of $\delta t=100$, corresponding to 1\% strain (recall that the globally imposed shear rate is $\gammadot=10^{-4}$). A survey of the layer-resolved mean squared displacements, $\text{MSD}_\parallel(\delta t)=\langle (\vec r_\parallel(t+\delta t)-\vec r_\parallel(t))^2\rangle$ (Fig.~\ref{fig:strAmpl}b) reveals that, at such a time scale, particles are trapped in their nearest neighbor cage and explore a region, whose size decreases closer to the wall. This connection between the single particle dynamic on intermediate times and the non-affine strain amplitude is underlined in Fig.~\ref{fig:strAmpl}a, where MSD$_\parallel(\delta t =100)$, normalized by the reference bulk value, is shown versus distance from the wall.

\section{Summary and conclusion}
In this letter, we study, via molecular dynamics simulations, wall effects on the spatial correlations of \blue{non-affine} strain under steady shear. Normalized spatial correlations of non-affine strain decay more slowly in the vicinity of a hard wall as compared to the bulk. In contrast to this, the amplitude of correlations, the square of the local non-affine strain, is strongly suppressed by the wall. The slower decay of correlations is interpreted as indicative of an enhancement of the elastic propagator in the proximity of a hard wall. This interpretation is born out with continuum mechanics calculations of the strain field around a pre-sheared spherical inclusion and validated further by molecular dynamics simulations of the same problem. The decrease of non-affine strain amplitude, on the other hand, reflects the single particle dynamics in a small region (the nearest neighbor cage) which tightens in the proximity of a hard wall with a particle scale roughness.

While the enhancement of correlations favors stronger \blue{strain fluctuations}, the suppression of the correlation amplitude acts against it. The overall effect of a rough wall in an amorphous solid will thus result from a competition of these opposing trends. This has an important implication for a mesoscopic modeling of deformation in amorphous solids in the presence of hard boundaries~\cite{Picard2005,Nicolas2013} as it states that accounting for wall effects on the elastic propagator alone is not sufficient. Rather, this information must be combined with a description of the spatial dependence of the strain amplitude from the distance to the wall. Since the strain amplitude reflects the particle dynamics on intermediate times, it will in general depend on the details of the wall-glass interactions. A thorough study of this issue would be an interesting topic for a future work.

\section{Acknowledgments}
M.~H.~ is supported by the German Research Foundation (DFG) under the project number VA 205/18-1. 
ICAMS acknowledges funding from its industrial sponsors, the state of North-Rhine Westphalia and the European Commission in the framework of the European Regional Development Fund (ERDF). Computation time by the J\"ulich supercomputing centre (ESMI 17) is acknowledged.

%\bibliographystyle{prsty}
%\bibliographystyle{eplbib}
%\bibliography{literature-2015-04-27}
\end{document}